\documentclass[aps,prl,reprint]{revtex4-1}
\usepackage[english]{babel}
\usepackage{graphicx}
\usepackage{graphics}
\usepackage{subfigure}
\usepackage{siunitx}
\usepackage{amsmath}
\begin{document}

\title{Material decomposition from a single x-ray projection
via single-grid phase contrast imaging}

\author{Celebrity F. Groenendijk\textsuperscript{1,2}}
\author{Florian Schaff\textsuperscript{2}}
\author{Linda C. P. Croton\textsuperscript{2}}
\author{Marcus J. Kitchen\textsuperscript{2}}
\author{Kaye S. Morgan\textsuperscript{2}}

\affiliation{\textsuperscript{1} Department of Radiation Science and Technology, Faculty of Applied Sciences, Delft University of Technology, Delft, The Netherlands}
\affiliation{\textsuperscript{2} School of Physics and Astronomy, Monash University, Clayton, Victoria, Australia}




\begin{abstract}
This study describes a new approach for material decomposition in x-ray imaging, utilising phase contrast to both increase sensitivity to weakly-attenuating samples and to act as a complementary measurement to attenuation, therefore allowing two overlaid materials to be separated.
The measurements are captured using the single-exposure, single-grid x-ray phase contrast imaging technique, with a novel correction that aims to remove propagation-based phase effects seen at sharp edges in the attenuation image.  The use of a single-exposure technique means that images could be collected in a high-speed sequence. Results are shown for both a known two-material sample and for a biological specimen.
\end{abstract}


\maketitle

\section{Introduction}

X-ray imaging is widely used as a non-invasive method of imaging the internal structure of a sample.  One difficulty in projection radiography is the inability to separate overlying features that are projected onto a two-dimensional image.  A recent clinical approach to address this has been dual-energy systems, which capture attenuation images at two different x-ray energies or spectra in order to separate out two composite materials from the anatomy \cite{Alvarez1976a}. A second difficulty in radiography is the inability to capture weakly-attenuating features, like the soft biological tissues that surround the strongly-attenuating bones.  Emerging techniques that capture x-ray phase effects have addressed this challenge \cite{Bravin2013}.  In this study, we utilise two simultaneous x-ray measurements, phase and attenuation, to separate out two materials from a single projection radiograph.  This approach provides the material separation advantage of dual-energy imaging with the soft-tissue sensitivity of phase contrast x-ray imaging. 

A range of experimental techniques have been developed for phase contrast x-ray imaging (PCXI), both at synchrotron x-ray sources and at laboratory-based sources \cite{Wilkins2014}.  In the case where a time-series of projection images is desired, the propagation-based PCXI technique is commonly used, where an extra distance (centimetres-to-metres) is introduced between the sample and detector to allow the wavefield to self-interfere. The resulting image contains phase and attenuation effects, and can be utilised to extract sample thickness in the case of a single-material sample \cite{paganin2002}. If phase and attenuation effects are to be separated, then an analyser-crystal, grating interferometer, edge-illumination or speckle-tracking system is typically used and multiple exposures captured at different optical settings \cite{endrizzi2018}. If the extracted phase and attenuation images need not be reconstructed at the detector, an alternative is to directly resolve a grid or speckle pattern and track distortions in this pattern to measure phase effects \cite{Wen2010, Morgan2012, Berujon2012}.  This `single-grid' approach has the advantage that only one exposure of the sample is required.

\begin{figure}[htbp]
    \centering
    \includegraphics[width=1\linewidth]{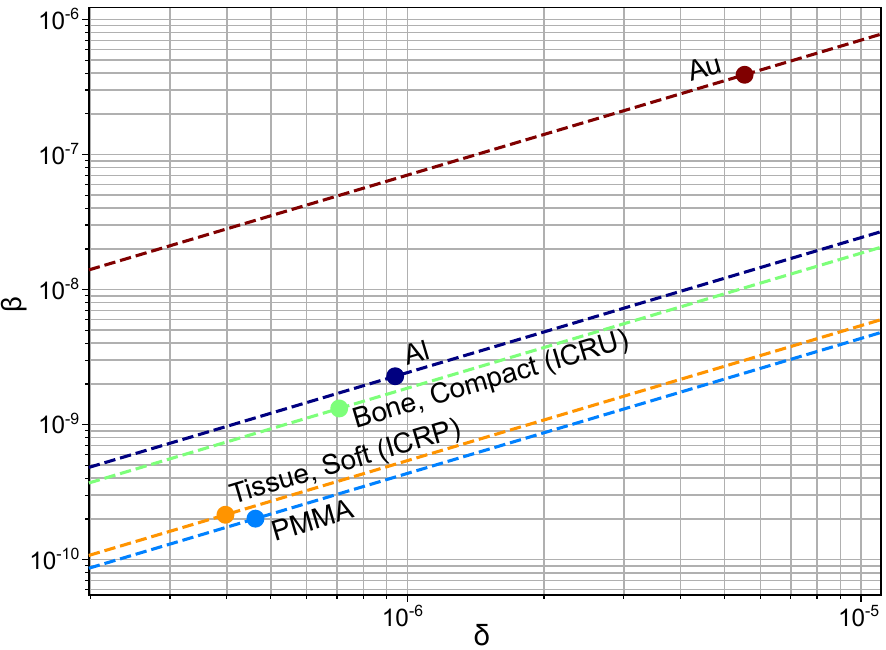}
    \caption{The $\delta$ and $\beta$ values for different materials at \SI{24}{\kilo\electronvolt} are shown as filled dots. A change in density with unchanged chemical composition leads to values along the dashed lines.}
    \label{fig:deltabetaplot}
\end{figure}

Material decomposition (MD) in X-ray imaging is the task of separating X-ray data, either two- or three-dimensional, into two or more of the sample's constituent materials. MD is commonly realised utilising the element-specific energy dependence of X-ray attenuation \cite{Alvarez1976a}, quantified by the linear attenuation coefficient $\mu$ of the element. It is, therefore, closely related to spectral X-ray imaging, which has been adopted to clinical imaging in the last decade \cite{McCollough2015}. X-ray phase-contrast imaging methods recover not only the x-ray attenuation, described by $\mu$, but also the phase shift imprinted by the sample, quantified by the refractive index decrement $\delta$ for a given element. As the ratio of $\delta$ and $\beta$ is element specific, spectral information can be substituted with attenuation and phase information for an alternative approach to MD. MD via phase contrast has so far been demonstrated experimentally using multiple-distance or energy propagation-based PCXI \cite{gureyev2002}, using a grating interferometer in combination with CT \cite{Qi2010b, Braig2018} and in projection using a crystal-analyser \cite{kitchen2011}. Figure \ref{fig:deltabetaplot} shows $\delta$ and $\beta$ for several materials at an X-ray energy of \SI{24}{\kilo\electronvolt}, with the ratio $\delta/\beta$ remaining constant under a change of sample density (dashed lines). For most robust decomposition, the $\delta/\beta$ ratio for the two composite materials should be as different as possible to each other. 

In this paper, we present MD using a single x-ray projection via single-grid phase contrast imaging \cite{morgan2011quantitative}. The single-grid technique recovers both attenuation and phase information and therefore allows for MD from a single exposure. We furthermore describe a novel correction method that removes propagation-based phase effects seen at sharp edges in the attenuation image. Experimental results of a known two-material sample and a biological specimen are presented.

\section{Imaging methods and results}
\label{sec:image_decomposition}
The single-grid x-ray phase contrast imaging technique uses a single exposure to measure two qualities of the projected sample: attenuation and phase information. From this information, we can extract two unknowns, the thicknesses of composite materials 1 and 2, provided $\delta$ and $\beta$ for each material are known.

\begin{figure}[h!]
    \setlength{\lineskip}{0.0pt}
    \centering
    \subfigure{\includegraphics[width=0.48\linewidth]{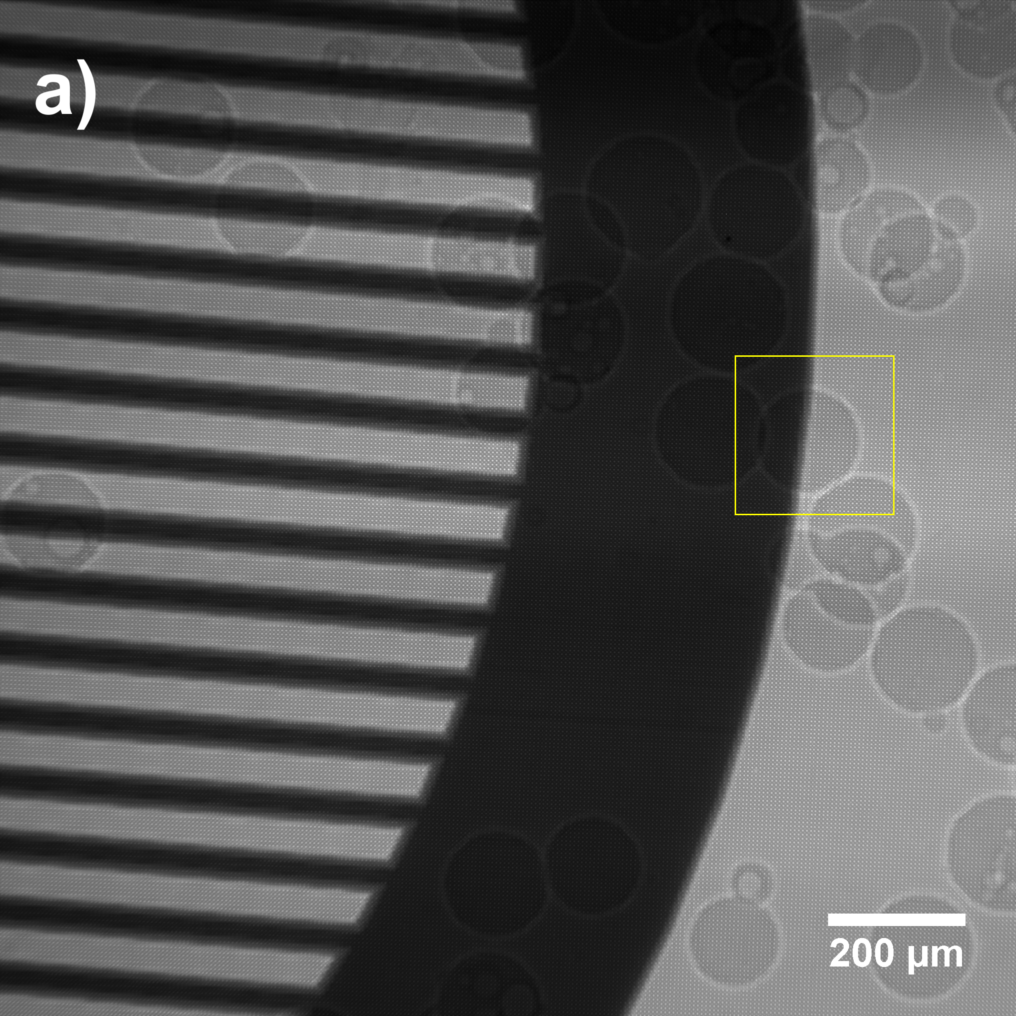} \label{fig:raw}}%
    \subfigure{\includegraphics[width=0.48\linewidth]{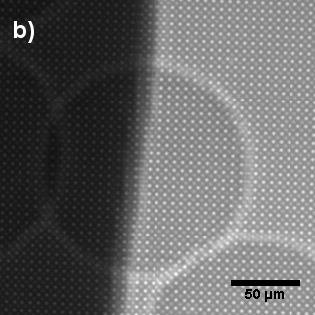} \label{fig:raw_magn}}\\
    \vspace*{-0.8em}
    \subfigure{\includegraphics[width=0.48\linewidth]{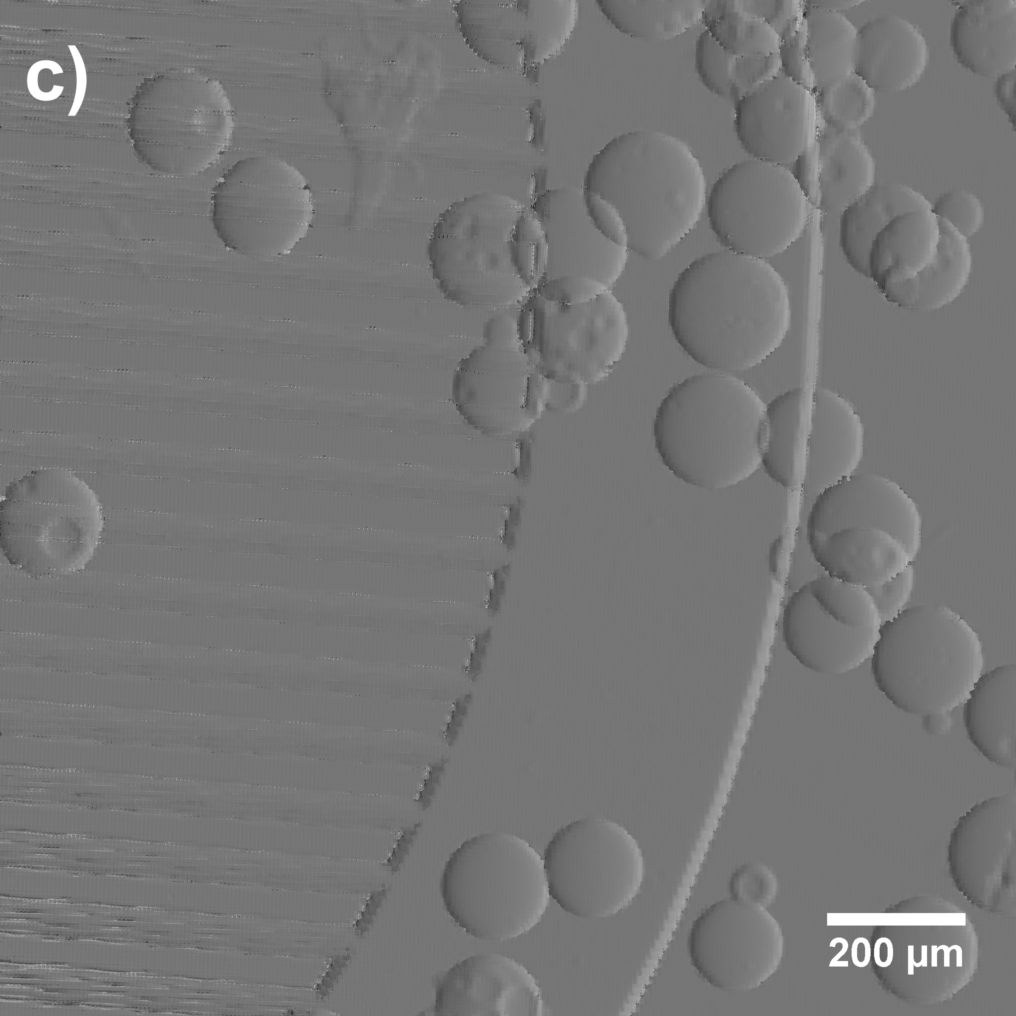} \label{fig:hori_phase}}%
    \subfigure{\includegraphics[width=0.48\linewidth]{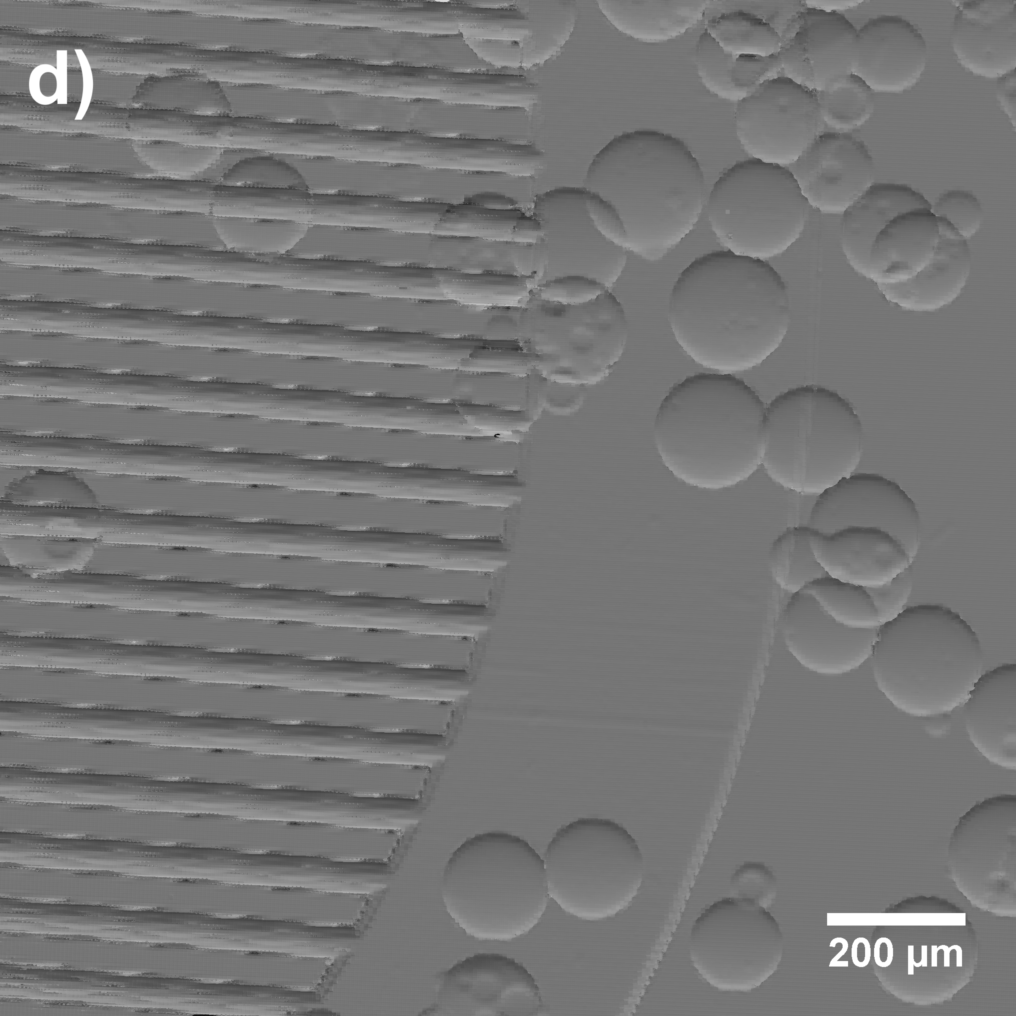} \label{fig:vert_phase}}
    \caption{a) Raw image containing both the reference grid pattern from the phase grid and the sample (gold EM grid and PMMA microspheres), also shown b) magnified to reveal the reference grid pattern. c) Horizontal and d) vertical differential phase images, shown here with a linear greyscale, stretching from -2 pixels shift (white) to +2 pixels (black).}
    \label{fig:raw_and_gradient}
\end{figure}

The experimental set-up used here places a phase grid between the x-ray source and the detector \cite{morgan2013phasegrid}. The grid provides a reference intensity pattern in order to acquire phase information. The presence of a sample - a gold bar grid (typically used in electron microscopy) and PMMA microspheres in our first example - causes a shift in the reference pattern, resulting from phase gradients present in the samples. Propagation of the phase-shifted wavefield over distance $z$, from the sample to the detector, locally distorts the reference pattern. From these local distortions or shifts in pattern, we are able to retrieve the phase depth. Figure \ref{fig:raw_and_gradient} shows the raw image (\ref{fig:raw}) containing both the reference grid pattern and the sample, with a magnified section shown in Fig. \ref{fig:raw_magn} to reveal the reference grid pattern.
The measurement of the local shift uses a modified version of the sub-pixel image registration algorithm of Guizar et al. \cite{guizar2008efficient}. At each pixel, the algorithm calculates the sub-pixel offset in the horizontal and vertical directions of the distorted grid pattern with respect to the reference image, based on a cross-correlation in Fourier space. By the use of the Fast Fourier Transform, an initial guess of the cross-correlation peak is made. A refinement of the peak location is accomplished by upsampling the Discrete Fourier Transform (DFT) within the vicinity of that peak location through a matrix-multiply DFT \cite{guizar2008efficient}. The output consists of a horizontal phase gradient ($S_x$) and a vertical phase gradient ($S_y$), shown in Fig. \ref{fig:hori_phase} and Fig. \ref{fig:vert_phase}, respectively. 

The samples were imaged with a \SI{5.4}{\um} period checkerboard phase grid 36.45 cm upstream of the detector, \SI{0.722}{\um} pixel size, \SI{25}{keV} synchrotron x-rays, 150 ms exposures and \SI{17}{cm} sample-to-detector distance. The phase retrieval via cross-correlation was performed at every pixel, investigating shifts of up to 2 pixels using a 8 $\times$ 8 pixel interrogation window. Subsequently, the transverse shifts of the grid reference pattern are converted into angles that describe the deflection of the wavefield over distance $z$: $\text{tan}(\theta_x)=S_x/z$ and $\text{tan}(\theta_y)=S_y/z$.

\begin{figure}[]
    \setlength{\lineskip}{2.0pt}
    \centering
    \subfigure{\includegraphics[width=0.48\linewidth]{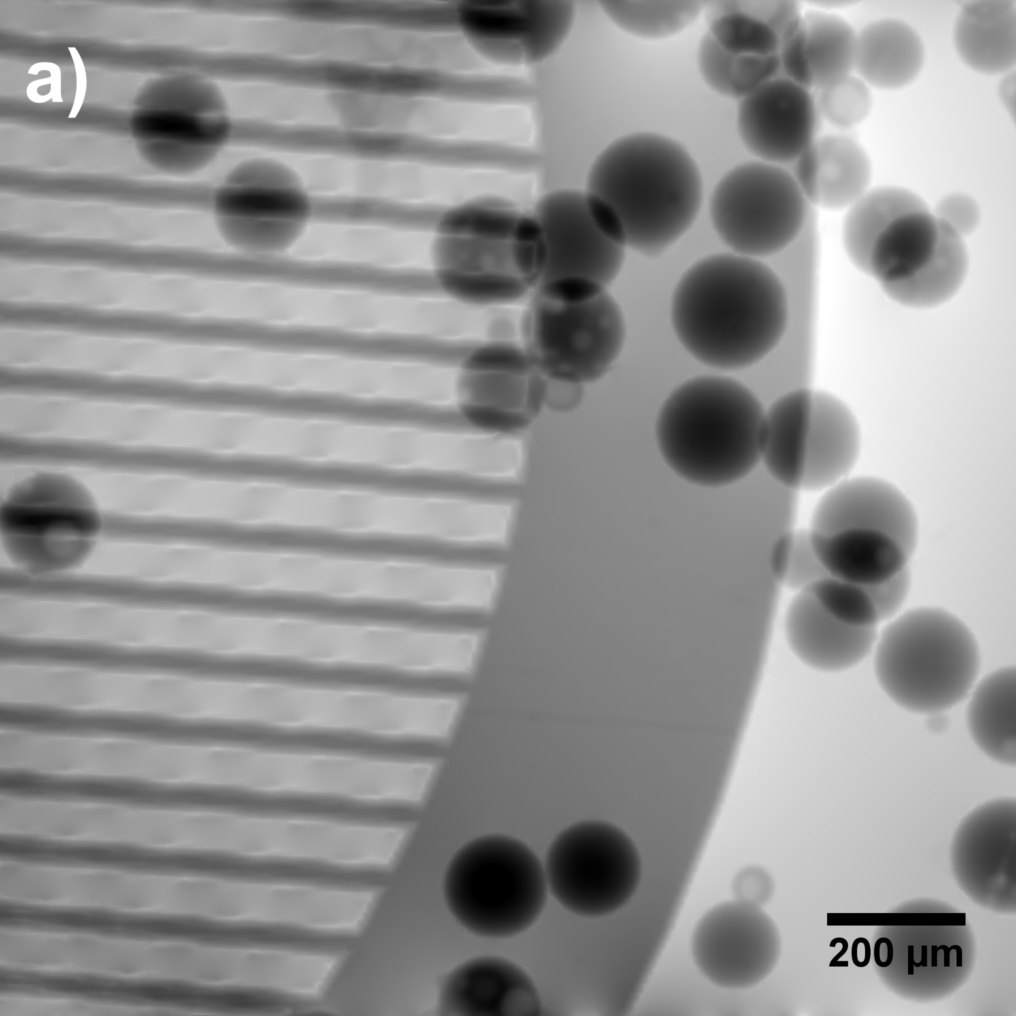} \label{fig:phasedepth}}%
    \subfigure{\includegraphics[width=0.48\linewidth]{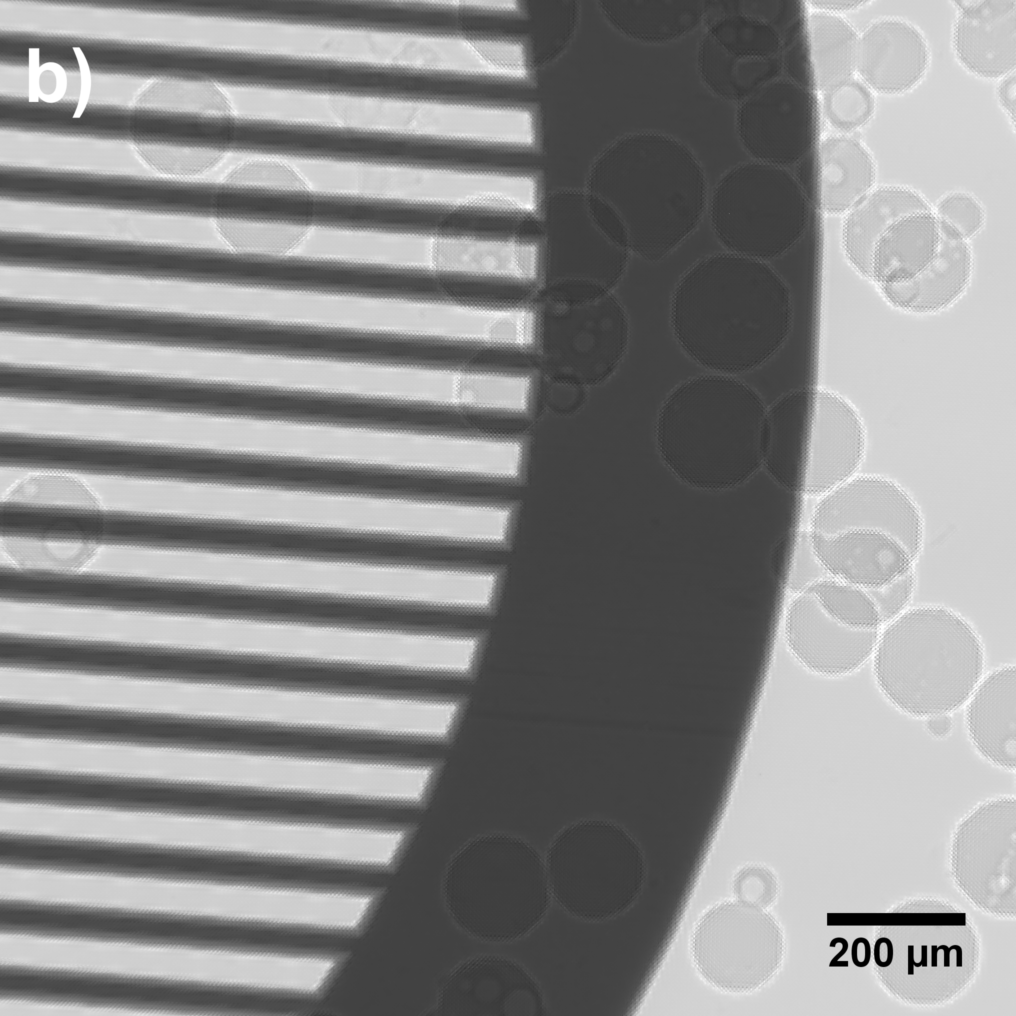} \label{fig:uncor_atten}}\\
    \vspace*{-0.8em}
    \subfigure{\includegraphics[width=0.48\linewidth]{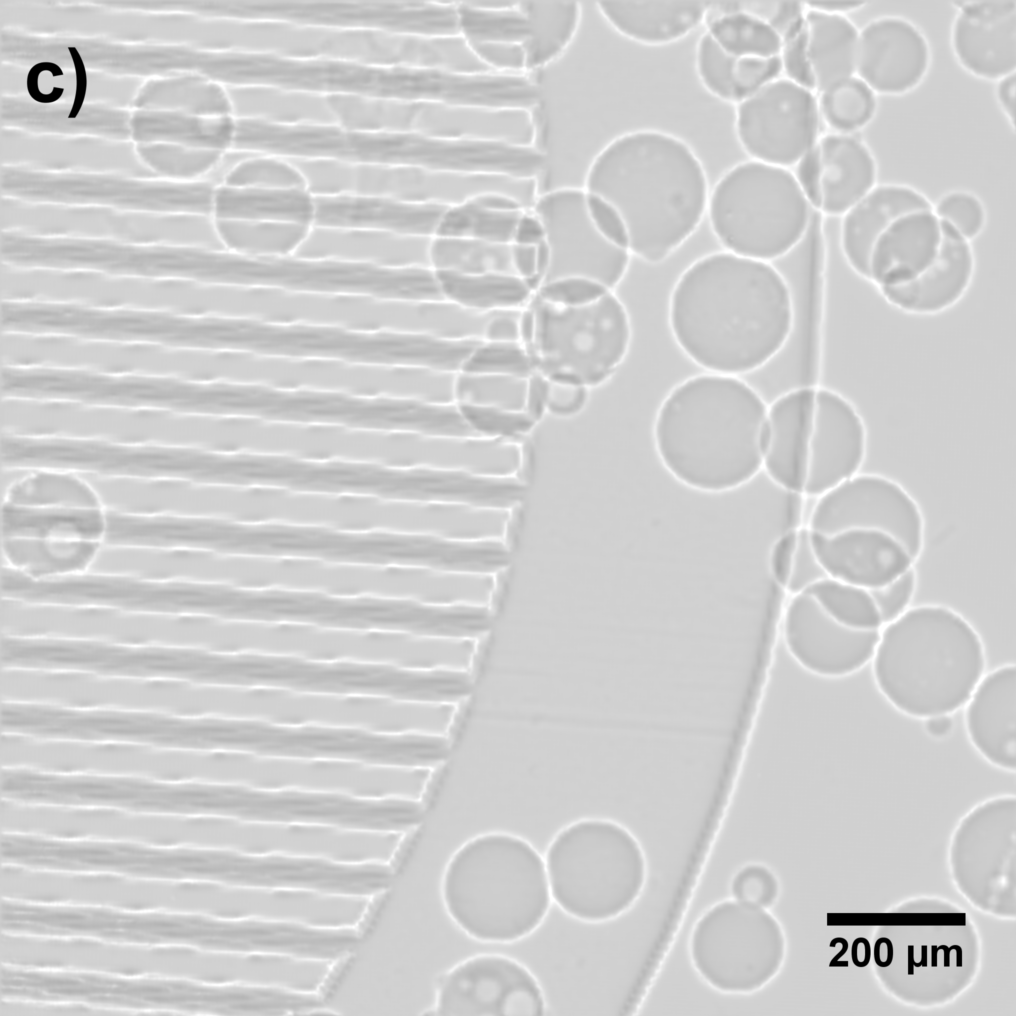} \label{fig:prop_phase}}%
    \subfigure{\includegraphics[width=0.48\linewidth]{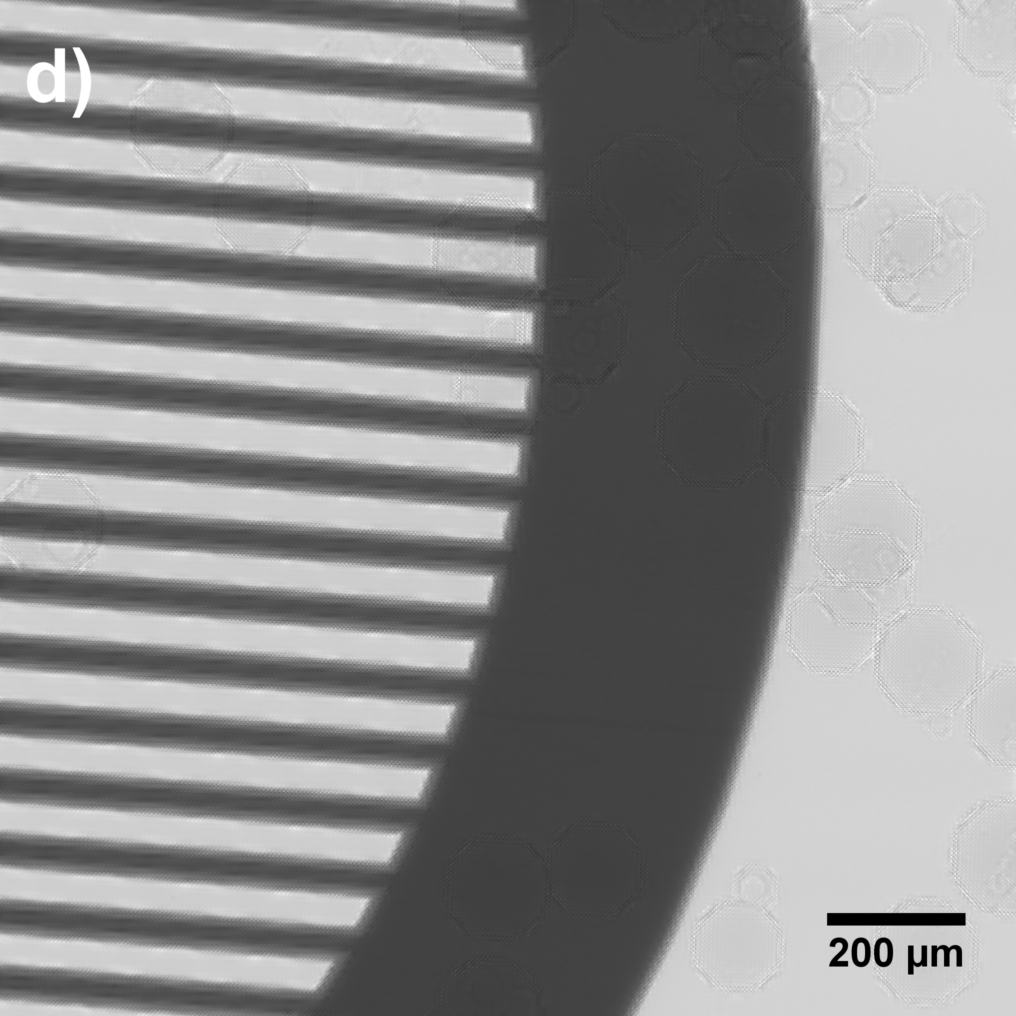} \label{fig:corr_atten}}
    \caption{For the gold/PMMA sample, a) the recovered phase depth image, b) the initial (uncorrected) attenuation image, c) the simulated image describing propagation-based phase effects, and d) the corrected attenuation image, where propagation-based phase effects are removed.}
    \label{fig:phase_and_atten}
\end{figure}

The sample's phase depth $\phi$ is described by $\phi=k \delta T$, with wavenumber $k=2\pi / \lambda \label{eq:phasedepth}$, refractive index decrement $\delta$, and the projected thickness $T$ \cite{paganin2006coherent}. The deflection of the incident rays with angles $\theta_x$ and $\theta_y$ follows the equations:
\begin{equation}
    \theta_x = \frac{1}{k} \frac{\partial \phi}{\partial x} \text{ and } \theta_y = \frac{1}{k} \frac{\partial \phi}{\partial y}.
\end{equation}
To retrieve the sample's phase depth, the phase gradient images are integrated according to the Fourier method following the formula:
\begin{equation}
    \phi (x,y) = \mathfrak{F}^{-1} \bigg[ \frac{\mathfrak{F} (\partial \phi / \partial x) + i \mathfrak{F} (\partial \phi / \partial y)}{ik_x-k_y} \bigg],
\end{equation}
where $k_x$ and $k_y$ are the Fourier coordinates and 
\begin{equation}
    \frac{\partial \phi}{\partial x} = k\tan^{-1}(S_x/z) \text{ and }
    \frac{\partial \phi}{\partial y} = k\tan^{-1}(S_y/z) \text{ \cite{morgan2011quantitative}.}
\end{equation}
The resultant phase depth is shown in Fig. \ref{fig:phasedepth}.

The image describing x-ray attenuation is formed by dividing the raw image (\ref{fig:raw}) by the grid only image (taken earlier) at the positions of the shifted grid pattern found by the sub-pixel registration algorithm. The attenuation image is shown in Fig. \ref{fig:uncor_atten}.

Since the synchrotron x-ray source is highly coherent and the pixel size is relatively small, even with a small distance between the sample and detector, the attenuation image (Fig. \ref{fig:uncor_atten}) includes propagation-based phase effects from the sharp edges within the sample. In particular, the PMMA spheres are each surrounded by a bright fringe that would be incorrectly interpreted in a purely attenuation-based image as `negative' attenuation. Given the reconstructed phase depth (Fig. \ref{fig:phasedepth}), it is possible to determine how we expect these fringes to appear. This was done by numerically propagating a wavefield with uniform intensity and with phase equal to the measured phase depth (Fig. \ref{fig:phasedepth}), using the angular spectrum approach \cite{als2011}, then blurring the resulting image by the measured point spread function. The propagation distance in this numerical propagation was set to the sample-to-detector distance used in the experiment, in this case \SI{17}{cm}. The resulting image, Fig. \ref{fig:prop_phase}, shows just the propagation-based phase effects. The attenuation image shown in Fig. \ref{fig:uncor_atten} is then divided by the numerically-predicted phase effects shown in Fig. \ref{fig:prop_phase} to produce Fig. \ref{fig:corr_atten}, the attenuation image that has been corrected to remove phase effects. Figures \ref{fig:phasedepth} and \ref{fig:corr_atten} can then be used as two independent measures of the sample to achieve MD.

The phase depth measurement depends on known values $k$ and $\delta$ and unknown thickness $T$ (equation \ref{eq:phasedepth}). A thickness map can be recovered by dividing the phase depth image by $k\delta$. In the case of two materials, the phase depth image is described as 
\begin{equation}
\phi = k\delta_1 T_1 + k\delta_2 T_2 = M_{\phi}, \label{eq:phaseimage}
\end{equation}
where $\phi$ indicates the phase depth image, with this measurement from now on written as $M_{\phi}$, and $T_1$ and $T_2$ are defined as the thicknesses of material 1 (gold) and material 2 (PMMA), respectively.

The attenuation image depends on the linear attenuation coefficient $\mu$ of a material, described by $I = I_0 e^{-\mu T}$ where $I$ is the measured intensity behind thickness $T$ of the material, $I_0$ is the original intensity at $T=0$ and $\mu$ is the linear attenuation coefficient. This equation holds for a single material. In the case of two materials, the equation becomes $I = I_0 e^{-\mu_1T_1 - \mu_2T_2}$.

The attenuation image is given by the fraction of transmitted photons $I$ with respect to the original intensity $I_0$, $I / I_0$. In the case of a single material, a thickness map can be recovered by taking the negative $\log$ of the attenuation image and dividing by $\mu$. In the case of two materials, the attenuation image follows the equation
\begin{equation}
-\ln(I/I_0) = \mu_1T_1 + \mu_2T_2 = M_{\mu},
\label{eq:attenimage}
\end{equation}
where $-\ln(I/I_0)$ is the negative logarithm of the attenuation image, with the measurement from now on written as $M_{\mu}$. 

Combining equations \ref{eq:phaseimage} and \ref{eq:attenimage} in matrix notation gives
\begin{equation}
    \begin{bmatrix} 
    k\delta_1 & k\delta_2 \\
    \mu_1 & \mu_2
    \end{bmatrix}
    \begin{bmatrix}
    T_1\\
    T_2
    \end{bmatrix}
    =
    \begin{bmatrix}
    M_{\phi}\\
    M_{\mu}
    \end{bmatrix},
\end{equation}
containing the known $k\delta\mu$-matrix, unknown thickness array, and known image array containing the phase depth image and attenuation image. The retrieval of the projected thickness map of each material can be obtained by taking the inverse of the $k\delta\mu$-matrix and multiplying with the image array. 

The formation of two thickness maps is based on the tensor dot product between the $k\delta\mu$-matrix and the image array, yielding the equations:
\begin{equation}
    T_1 = \frac{1}{k\delta_1\mu_2 - k\delta_2\mu_1} \cdot (\mu_2M_\phi - k\delta_2M_\mu) \label{eq:thickness1}
\end{equation}
and
\begin{equation}
    T_2 = \frac{1}{k\delta_1\mu_2 - k\delta_2\mu_1} \cdot (-\mu_1M_\phi + k\delta_1M_\mu). \label{eq:thickness2}
\end{equation}

\begin{figure}[htbp]
    \centering
    \subfigure{\includegraphics[width=0.48\linewidth]{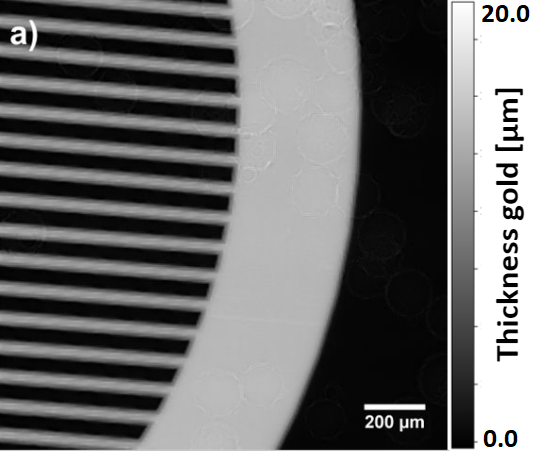} \label{fig:T_gold}}
    \subfigure{\includegraphics[width=0.48\linewidth]{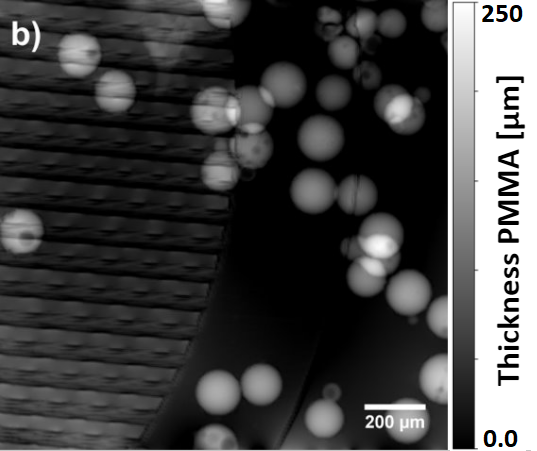} \label{fig:T_pmma}}
    \caption{Thickness maps ($T_1,T_2$) of gold (a) and PMMA (b) as a result of MD with accompanying colorbar describing the extracted thickness values in micrometer.}
    \label{fig:thicknesses}
\end{figure}

The result of MD is presented in Fig. \ref{fig:thicknesses} where Fig. \ref{fig:T_gold} shows the extracted thickness map of gold and Fig. \ref{fig:T_pmma} shows the thickness map of the PMMA microspheres. The scalebar/greyscale bar corresponds to the extracted thickness of the material in micrometres. 

\begin{figure}[htbp]
    \centering
    \subfigure{\includegraphics[width=0.32\linewidth]{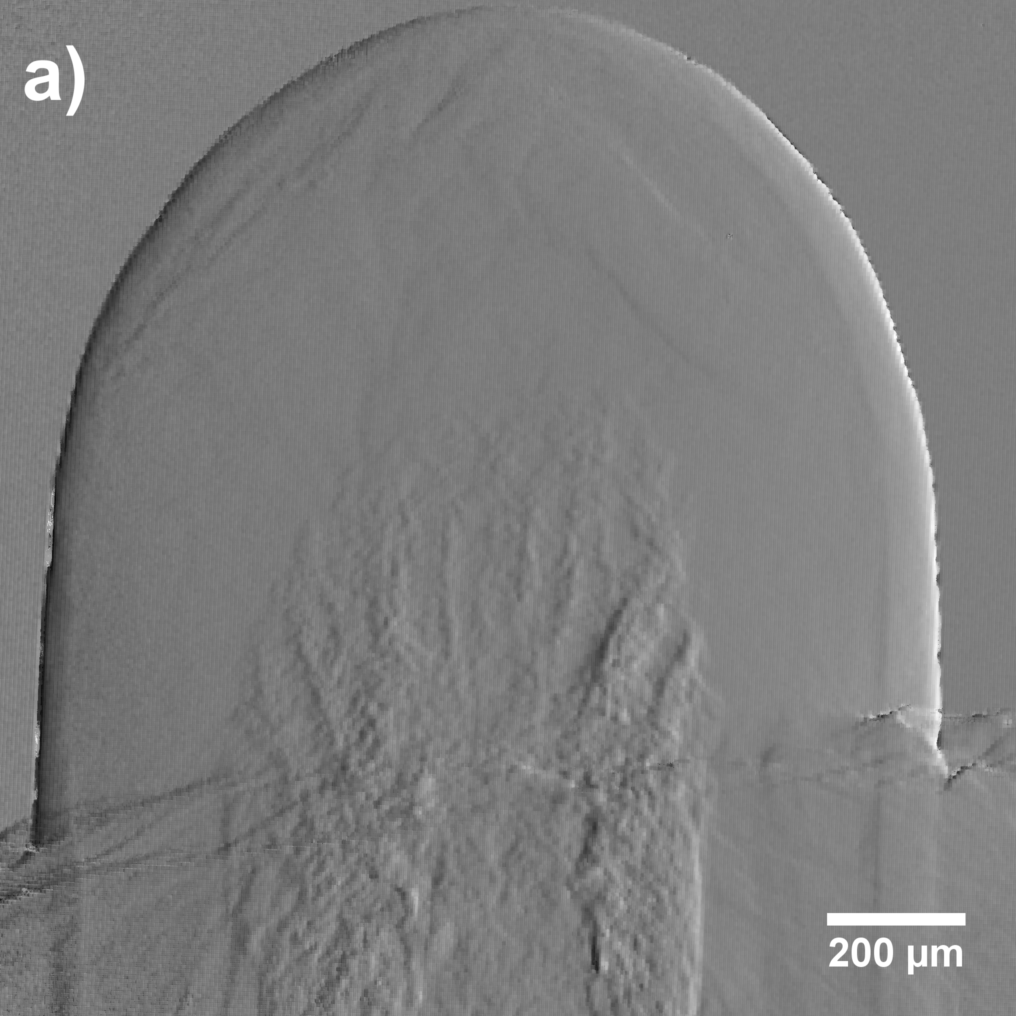}\label{fig:toe_phasedepth}}
    \subfigure{\includegraphics[width=0.32\linewidth]{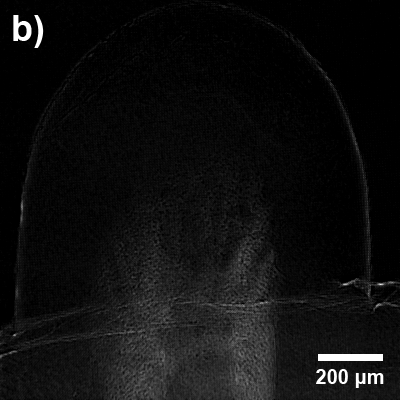}\label{fig:toe_atten}}
    \subfigure{\includegraphics[width=0.32\linewidth]{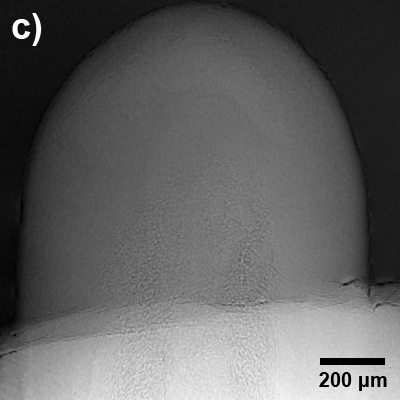}\label{fig:toe_thickness}}
    \caption{PCXI images of an immature rabbit toe. The horizontal phase gradient image (a), the thickness map of bone (b) and the thickness map of tissue (c).}
    \label{fig:toe_images}
\end{figure}

To test the algorithm on a biological sample, a rabbit toe was imaged using the same experimental parameters, with thicknesses reconstructed using the $\delta$ and $\mu$ values for bone and soft tissue. The results can be seen in Fig. \ref{fig:toe_images}, including the horizontal phase gradient (Fig. \ref{fig:toe_phasedepth}), then the extracted thickness of the bone (Fig. \ref{fig:toe_atten}) and the soft tissue (Fig. \ref{fig:toe_thickness}).

\section{Discussion}
We have demonstrated material decomposition from a single image at a single x-ray energy using the single-grid X-ray phase-contrast imaging technique. Most commonly, phase MD has been performed using phase imaging techniques that utilise one-dimensional gratings or an analyser-crystal and hence capture just one direction of the phase gradient \cite{Qi2010b,Braig2018, kitchen2011}. This makes phase integration highly susceptible to noise, and hence MD via phase contrast has been most successful in practice using three-dimensional computed tomography (CT). In contrast, the single-grid method presented here allows for accurate integration of the phase from a single image, due to simultaneous measurements of the phase gradients in both image directions \cite{morgan2016}. Consequently, MD can be performed using a single two-dimensional projection image, which is beneficial, for example, in rapid imaging aiming to capture dynamics. The translation of the single-grid method to CT is straightforward; and the methods developed in \cite{Qi2010b} and \cite{Braig2018} can readily be applied after obtaining the phase and attenuation reconstructions, or CT reconstruction could be performed using the material thickness maps (e.g. Fig. \ref{fig:T_gold} and \ref{fig:T_pmma}) at each projection angle.

The key limitation of MD is that the sample must contain no more than two other materials in addition to air. If additional composite materials need to be extracted, then additional complementary measurements are required, potentially at multiple energies (e.g with an energy-resolving detector).  Further work could investigate areas where the separation does not work perfectly, for example the periodic artefacts seen where the reference grid intercepts the almost horizontal gold bars.  The imperfections seen in the decomposition of our dataset seem to arise from an imperfect phase integration that contains some low-frequency artefacts.

The phase propagation method used to correct residual phase-contrast fringes in the attenuation images (shown in Fig. \ref{fig:phase_and_atten}) has the potential to improve any kind of differential phase imaging that suffers from propagation-based edge fringes. The method presented here serves as a way to substantially reduce these effects that can otherwise reduce the accuracy of quantitative measurements or result in unwanted image artefacts. The results also show that, in cases like the imaging of the PMMA spheres, phase effects can play a significant role in reducing the intensity seen behind the sample features.

There are possible applications for this method in security, medicine and manufacturing, providing the most significant advantage in cases that require high-speed projection imaging of overlaid materials. In medicine, one example would be separating the ribs from the lungs in chest imaging, to accurately measure lung volume, which is already relevant in biomedical research into the aeration of lungs at birth \cite{kitchen2008}. A key problem in airport security screening is separating overlapping objects in a projection image of a suitcase \cite{zentai2008}, currently addressed using dual energy imaging, which lacks sensitivity to weakly-absorbing structures. Recent work in security screening has seen the most benefit from the scatter/dark-field modality \cite{miller2013}, with differential phase images revealing structures but making it hard to differentiate specific objects.  We anticipate that the approach here may be useful for this (e.g. compare attenuation, Fig. \ref{fig:corr_atten}, with a single-material reconstruction, Fig. \ref{fig:T_pmma}) and other such applications.

\section{Funding Information}

Australian Research Council Future Fellowships FT180100374, FT160100454; Australian Research Council Discovery Project DP170103678; SPring-8 Beamtime 2018B1315 and 2019A1151 on BL20XU, with the approval of the Japan Synchrotron Radiation Research Institute (JASRI); Veski Victorian Postdoctoral Research Fellowship (VPRF); German Excellence Initiative and European Union Seventh Framework Program (291763).

We thank Martin Donnelley for his assistance with the experimental set-up, and Christian David, Simon Rutishauser and Vitaliy Guzenko for the fabrication of the phase grid.

\bibliography{References}

\begin{thebibliography}{22}%
\makeatletter
\providecommand \@ifxundefined [1]{%
 \@ifx{#1\undefined}
}%
\providecommand \@ifnum [1]{%
 \ifnum #1\expandafter \@firstoftwo
 \else \expandafter \@secondoftwo
 \fi
}%
\providecommand \@ifx [1]{%
 \ifx #1\expandafter \@firstoftwo
 \else \expandafter \@secondoftwo
 \fi
}%
\providecommand \natexlab [1]{#1}%
\providecommand \enquote  [1]{``#1''}%
\providecommand \bibnamefont  [1]{#1}%
\providecommand \bibfnamefont [1]{#1}%
\providecommand \citenamefont [1]{#1}%
\providecommand \href@noop [0]{\@secondoftwo}%
\providecommand \href [0]{\begingroup \@sanitize@url \@href}%
\providecommand \@href[1]{\@@startlink{#1}\@@href}%
\providecommand \@@href[1]{\endgroup#1\@@endlink}%
\providecommand \@sanitize@url [0]{\catcode `\\12\catcode `\$12\catcode
  `\&12\catcode `\#12\catcode `\^12\catcode `\_12\catcode `\%12\relax}%
\providecommand \@@startlink[1]{}%
\providecommand \@@endlink[0]{}%
\providecommand \url  [0]{\begingroup\@sanitize@url \@url }%
\providecommand \@url [1]{\endgroup\@href {#1}{\urlprefix }}%
\providecommand \urlprefix  [0]{URL }%
\providecommand \Eprint [0]{\href }%
\providecommand \doibase [0]{http://dx.doi.org/}%
\providecommand \selectlanguage [0]{\@gobble}%
\providecommand \bibinfo  [0]{\@secondoftwo}%
\providecommand \bibfield  [0]{\@secondoftwo}%
\providecommand \translation [1]{[#1]}%
\providecommand \BibitemOpen [0]{}%
\providecommand \bibitemStop [0]{}%
\providecommand \bibitemNoStop [0]{.\EOS\space}%
\providecommand \EOS [0]{\spacefactor3000\relax}%
\providecommand \BibitemShut  [1]{\csname bibitem#1\endcsname}%
\let\auto@bib@innerbib\@empty
\bibitem [{\citenamefont {Alvarez}\ and\ \citenamefont
  {Macovski}(1976)}]{Alvarez1976a}%
  \BibitemOpen
  \bibfield  {author} {\bibinfo {author} {\bibfnamefont {R.~E.}\ \bibnamefont
  {Alvarez}}\ and\ \bibinfo {author} {\bibfnamefont {A.}~\bibnamefont
  {Macovski}},\ }\href {\doibase 10.1088/0031-9155/21/5/002} {\bibfield
  {journal} {\bibinfo  {journal} {Phys. Med. Biol.}\ }\textbf {\bibinfo
  {volume} {21}},\ \bibinfo {pages} {002} (\bibinfo {year} {1976})}\BibitemShut
  {NoStop}%
\bibitem [{\citenamefont {Bravin}\ \emph {et~al.}(2013)\citenamefont {Bravin},
  \citenamefont {Coan},\ and\ \citenamefont {Suortti}}]{Bravin2013}%
  \BibitemOpen
  \bibfield  {author} {\bibinfo {author} {\bibfnamefont {A.}~\bibnamefont
  {Bravin}}, \bibinfo {author} {\bibfnamefont {P.}~\bibnamefont {Coan}}, \ and\
  \bibinfo {author} {\bibfnamefont {P.}~\bibnamefont {Suortti}},\ }\href@noop
  {} {\bibfield  {journal} {\bibinfo  {journal} {Physics in Medicine and
  Biology}\ }\textbf {\bibinfo {volume} {58}},\ \bibinfo {pages} {R1} (\bibinfo
  {year} {2013})}\BibitemShut {NoStop}%
\bibitem [{\citenamefont {Wilkins}\ \emph {et~al.}(2014)\citenamefont
  {Wilkins}, \citenamefont {Nesterets}, \citenamefont {Gureyev}, \citenamefont
  {Mayo}, \citenamefont {Pogany},\ and\ \citenamefont
  {Stevenson}}]{Wilkins2014}%
  \BibitemOpen
  \bibfield  {author} {\bibinfo {author} {\bibfnamefont {S.}~\bibnamefont
  {Wilkins}}, \bibinfo {author} {\bibfnamefont {Y.~I.}\ \bibnamefont
  {Nesterets}}, \bibinfo {author} {\bibfnamefont {T.}~\bibnamefont {Gureyev}},
  \bibinfo {author} {\bibfnamefont {S.}~\bibnamefont {Mayo}}, \bibinfo {author}
  {\bibfnamefont {A.}~\bibnamefont {Pogany}}, \ and\ \bibinfo {author}
  {\bibfnamefont {A.}~\bibnamefont {Stevenson}},\ }\href@noop {} {\bibfield
  {journal} {\bibinfo  {journal} {Philosophical Transactions of the Royal
  Society A}\ }\textbf {\bibinfo {volume} {372}},\ \bibinfo {pages} {20130021}
  (\bibinfo {year} {2014})}\BibitemShut {NoStop}%
\bibitem [{\citenamefont {Paganin}\ \emph {et~al.}(2002)\citenamefont
  {Paganin}, \citenamefont {Mayo}, \citenamefont {Gureyev}, \citenamefont
  {Miller},\ and\ \citenamefont {Wilkins}}]{paganin2002}%
  \BibitemOpen
  \bibfield  {author} {\bibinfo {author} {\bibfnamefont {D.}~\bibnamefont
  {Paganin}}, \bibinfo {author} {\bibfnamefont {S.~C.}\ \bibnamefont {Mayo}},
  \bibinfo {author} {\bibfnamefont {T.~E.}\ \bibnamefont {Gureyev}}, \bibinfo
  {author} {\bibfnamefont {P.~R.}\ \bibnamefont {Miller}}, \ and\ \bibinfo
  {author} {\bibfnamefont {S.~W.}\ \bibnamefont {Wilkins}},\ }\href@noop {}
  {\bibfield  {journal} {\bibinfo  {journal} {Journal of Microscopy}\ }\textbf
  {\bibinfo {volume} {206}},\ \bibinfo {pages} {33} (\bibinfo {year}
  {2002})}\BibitemShut {NoStop}%
\bibitem [{\citenamefont {Endrizzi}(2018)}]{endrizzi2018}%
  \BibitemOpen
  \bibfield  {author} {\bibinfo {author} {\bibfnamefont {M.}~\bibnamefont
  {Endrizzi}},\ }\href@noop {} {\bibfield  {journal} {\bibinfo  {journal}
  {Nuclear instruments and methods in physics research section A: Accelerators,
  spectrometers, detectors and associated equipment}\ }\textbf {\bibinfo
  {volume} {878}},\ \bibinfo {pages} {88} (\bibinfo {year} {2018})}\BibitemShut
  {NoStop}%
\bibitem [{\citenamefont {Wen}\ \emph {et~al.}(2010)\citenamefont {Wen},
  \citenamefont {Bennett}, \citenamefont {Kopace}, \citenamefont {Stein},\ and\
  \citenamefont {Pai}}]{Wen2010}%
  \BibitemOpen
  \bibfield  {author} {\bibinfo {author} {\bibfnamefont {H.~H.}\ \bibnamefont
  {Wen}}, \bibinfo {author} {\bibfnamefont {E.~E.}\ \bibnamefont {Bennett}},
  \bibinfo {author} {\bibfnamefont {R.}~\bibnamefont {Kopace}}, \bibinfo
  {author} {\bibfnamefont {A.~F.}\ \bibnamefont {Stein}}, \ and\ \bibinfo
  {author} {\bibfnamefont {V.}~\bibnamefont {Pai}},\ }\href@noop {} {\bibfield
  {journal} {\bibinfo  {journal} {Optics Letters}\ }\textbf {\bibinfo {volume}
  {35}},\ \bibinfo {pages} {1932} (\bibinfo {year} {2010})}\BibitemShut
  {NoStop}%
\bibitem [{\citenamefont {Morgan}\ \emph {et~al.}(2012)\citenamefont {Morgan},
  \citenamefont {Paganin},\ and\ \citenamefont {Siu}}]{Morgan2012}%
  \BibitemOpen
  \bibfield  {author} {\bibinfo {author} {\bibfnamefont {K.~S.}\ \bibnamefont
  {Morgan}}, \bibinfo {author} {\bibfnamefont {D.~M.}\ \bibnamefont {Paganin}},
  \ and\ \bibinfo {author} {\bibfnamefont {K.~K.~W.}\ \bibnamefont {Siu}},\
  }\href {\doibase 10.1063/1.3694918} {\bibfield  {journal} {\bibinfo
  {journal} {Applied Physics Letters}\ }\textbf {\bibinfo {volume} {100}},\
  \bibinfo {pages} {124102} (\bibinfo {year} {2012})}\BibitemShut {NoStop}%
\bibitem [{\citenamefont {B\'erujon}\ \emph {et~al.}(2012)\citenamefont
  {B\'erujon}, \citenamefont {Ziegler}, \citenamefont {Cerbino},\ and\
  \citenamefont {Peverini}}]{Berujon2012}%
  \BibitemOpen
  \bibfield  {author} {\bibinfo {author} {\bibfnamefont {S.}~\bibnamefont
  {B\'erujon}}, \bibinfo {author} {\bibfnamefont {E.}~\bibnamefont {Ziegler}},
  \bibinfo {author} {\bibfnamefont {R.}~\bibnamefont {Cerbino}}, \ and\
  \bibinfo {author} {\bibfnamefont {L.}~\bibnamefont {Peverini}},\ }\href
  {\doibase 10.1103/PhysRevLett.108.158102} {\bibfield  {journal} {\bibinfo
  {journal} {Physical Review Letters}\ }\textbf {\bibinfo {volume} {108}}
  (\bibinfo {year} {2012}),\ 10.1103/PhysRevLett.108.158102}\BibitemShut
  {NoStop}%
\bibitem [{\citenamefont {McCollough}\ \emph {et~al.}(2015)\citenamefont
  {McCollough}, \citenamefont {Leng}, \citenamefont {Yu},\ and\ \citenamefont
  {Fletcher}}]{McCollough2015}%
  \BibitemOpen
  \bibfield  {author} {\bibinfo {author} {\bibfnamefont {C.~H.}\ \bibnamefont
  {McCollough}}, \bibinfo {author} {\bibfnamefont {S.}~\bibnamefont {Leng}},
  \bibinfo {author} {\bibfnamefont {L.}~\bibnamefont {Yu}}, \ and\ \bibinfo
  {author} {\bibfnamefont {J.~G.}\ \bibnamefont {Fletcher}},\ }\href {\doibase
  10.1148/radiol.2015142631} {\bibfield  {journal} {\bibinfo  {journal}
  {Radiology}\ }\textbf {\bibinfo {volume} {276}},\ \bibinfo {pages} {637}
  (\bibinfo {year} {2015})}\BibitemShut {NoStop}%
\bibitem [{\citenamefont {Gureyev}\ \emph {et~al.}(2002)\citenamefont
  {Gureyev}, \citenamefont {Stevenson}, \citenamefont {Paganin}, \citenamefont
  {Weitkamp}, \citenamefont {Snigirev}, \citenamefont {Snigireva},\ and\
  \citenamefont {Wilkins}}]{gureyev2002}%
  \BibitemOpen
  \bibfield  {author} {\bibinfo {author} {\bibfnamefont {T.~E.}\ \bibnamefont
  {Gureyev}}, \bibinfo {author} {\bibfnamefont {A.~W.}\ \bibnamefont
  {Stevenson}}, \bibinfo {author} {\bibfnamefont {D.~M.}\ \bibnamefont
  {Paganin}}, \bibinfo {author} {\bibfnamefont {T.}~\bibnamefont {Weitkamp}},
  \bibinfo {author} {\bibfnamefont {A.}~\bibnamefont {Snigirev}}, \bibinfo
  {author} {\bibfnamefont {I.}~\bibnamefont {Snigireva}}, \ and\ \bibinfo
  {author} {\bibfnamefont {S.}~\bibnamefont {Wilkins}},\ }\href@noop {}
  {\bibfield  {journal} {\bibinfo  {journal} {Journal of Synchrotron
  Radiation}\ }\textbf {\bibinfo {volume} {9}},\ \bibinfo {pages} {148}
  (\bibinfo {year} {2002})}\BibitemShut {NoStop}%
\bibitem [{\citenamefont {Qi}\ \emph {et~al.}(2010)\citenamefont {Qi},
  \citenamefont {Zambelli}, \citenamefont {Bevins},\ and\ \citenamefont
  {Chen}}]{Qi2010b}%
  \BibitemOpen
  \bibfield  {author} {\bibinfo {author} {\bibfnamefont {Z.}~\bibnamefont
  {Qi}}, \bibinfo {author} {\bibfnamefont {J.}~\bibnamefont {Zambelli}},
  \bibinfo {author} {\bibfnamefont {N.}~\bibnamefont {Bevins}}, \ and\ \bibinfo
  {author} {\bibfnamefont {G.-H.}\ \bibnamefont {Chen}},\ }\href {\doibase
  10.1088/0031-9155/55/9/016} {\bibfield  {journal} {\bibinfo  {journal} {Phys.
  Med. Biol.}\ }\textbf {\bibinfo {volume} {55}},\ \bibinfo {pages} {2669}
  (\bibinfo {year} {2010})}\BibitemShut {NoStop}%
\bibitem [{\citenamefont {Braig}\ \emph {et~al.}(2018)\citenamefont {Braig},
  \citenamefont {B{\"{o}}hm}, \citenamefont {Dierolf}, \citenamefont {Jud},
  \citenamefont {G{\"{u}}nther}, \citenamefont {Mechlem}, \citenamefont
  {Allner}, \citenamefont {Sellerer}, \citenamefont {Achterhold}, \citenamefont
  {Gleich}, \citenamefont {No{\"{e}}l}, \citenamefont {Pfeiffer}, \citenamefont
  {Rummeny}, \citenamefont {Herzen},\ and\ \citenamefont
  {Pfeiffer}}]{Braig2018}%
  \BibitemOpen
  \bibfield  {author} {\bibinfo {author} {\bibfnamefont {E.}~\bibnamefont
  {Braig}}, \bibinfo {author} {\bibfnamefont {J.}~\bibnamefont {B{\"{o}}hm}},
  \bibinfo {author} {\bibfnamefont {M.}~\bibnamefont {Dierolf}}, \bibinfo
  {author} {\bibfnamefont {C.}~\bibnamefont {Jud}}, \bibinfo {author}
  {\bibfnamefont {B.}~\bibnamefont {G{\"{u}}nther}}, \bibinfo {author}
  {\bibfnamefont {K.}~\bibnamefont {Mechlem}}, \bibinfo {author} {\bibfnamefont
  {S.}~\bibnamefont {Allner}}, \bibinfo {author} {\bibfnamefont
  {T.}~\bibnamefont {Sellerer}}, \bibinfo {author} {\bibfnamefont
  {K.}~\bibnamefont {Achterhold}}, \bibinfo {author} {\bibfnamefont
  {B.}~\bibnamefont {Gleich}}, \bibinfo {author} {\bibfnamefont
  {P.}~\bibnamefont {No{\"{e}}l}}, \bibinfo {author} {\bibfnamefont
  {D.}~\bibnamefont {Pfeiffer}}, \bibinfo {author} {\bibfnamefont
  {E.}~\bibnamefont {Rummeny}}, \bibinfo {author} {\bibfnamefont
  {J.}~\bibnamefont {Herzen}}, \ and\ \bibinfo {author} {\bibfnamefont
  {F.}~\bibnamefont {Pfeiffer}},\ }\href {\doibase 10.1038/s41598-018-34809-6}
  {\bibfield  {journal} {\bibinfo  {journal} {Sci. Rep.}\ }\textbf {\bibinfo
  {volume} {8}},\ \bibinfo {pages} {16394} (\bibinfo {year}
  {2018})}\BibitemShut {NoStop}%
\bibitem [{\citenamefont {Kitchen}\ \emph {et~al.}(2011)\citenamefont
  {Kitchen}, \citenamefont {Paganin}, \citenamefont {Uesugi}, \citenamefont
  {Allison}, \citenamefont {Lewis}, \citenamefont {Hooper},\ and\ \citenamefont
  {Pavlov}}]{kitchen2011}%
  \BibitemOpen
  \bibfield  {author} {\bibinfo {author} {\bibfnamefont {M.~J.}\ \bibnamefont
  {Kitchen}}, \bibinfo {author} {\bibfnamefont {D.~M.}\ \bibnamefont
  {Paganin}}, \bibinfo {author} {\bibfnamefont {K.}~\bibnamefont {Uesugi}},
  \bibinfo {author} {\bibfnamefont {B.~J.}\ \bibnamefont {Allison}}, \bibinfo
  {author} {\bibfnamefont {R.~A.}\ \bibnamefont {Lewis}}, \bibinfo {author}
  {\bibfnamefont {S.~B.}\ \bibnamefont {Hooper}}, \ and\ \bibinfo {author}
  {\bibfnamefont {K.~M.}\ \bibnamefont {Pavlov}},\ }\href@noop {} {\bibfield
  {journal} {\bibinfo  {journal} {Physics in Medicine \& Biology}\ }\textbf
  {\bibinfo {volume} {56}},\ \bibinfo {pages} {515} (\bibinfo {year}
  {2011})}\BibitemShut {NoStop}%
\bibitem [{\citenamefont {Morgan}\ \emph {et~al.}(2011)\citenamefont {Morgan},
  \citenamefont {Paganin},\ and\ \citenamefont {Siu}}]{morgan2011quantitative}%
  \BibitemOpen
  \bibfield  {author} {\bibinfo {author} {\bibfnamefont {K.~S.}\ \bibnamefont
  {Morgan}}, \bibinfo {author} {\bibfnamefont {D.~M.}\ \bibnamefont {Paganin}},
  \ and\ \bibinfo {author} {\bibfnamefont {K.~K.}\ \bibnamefont {Siu}},\
  }\href@noop {} {\bibfield  {journal} {\bibinfo  {journal} {Optics Express}\
  }\textbf {\bibinfo {volume} {19}},\ \bibinfo {pages} {19781} (\bibinfo {year}
  {2011})}\BibitemShut {NoStop}%
\bibitem [{\citenamefont {Morgan}\ \emph {et~al.}(2013)\citenamefont {Morgan},
  \citenamefont {Modregger}, \citenamefont {Irvine}, \citenamefont
  {Rutishauser}, \citenamefont {Guzenko}, \citenamefont {Stampanoni},\ and\
  \citenamefont {David}}]{morgan2013phasegrid}%
  \BibitemOpen
  \bibfield  {author} {\bibinfo {author} {\bibfnamefont {K.~S.}\ \bibnamefont
  {Morgan}}, \bibinfo {author} {\bibfnamefont {P.}~\bibnamefont {Modregger}},
  \bibinfo {author} {\bibfnamefont {S.~C.}\ \bibnamefont {Irvine}}, \bibinfo
  {author} {\bibfnamefont {S.}~\bibnamefont {Rutishauser}}, \bibinfo {author}
  {\bibfnamefont {V.~A.}\ \bibnamefont {Guzenko}}, \bibinfo {author}
  {\bibfnamefont {M.}~\bibnamefont {Stampanoni}}, \ and\ \bibinfo {author}
  {\bibfnamefont {C.}~\bibnamefont {David}},\ }\href@noop {} {\bibfield
  {journal} {\bibinfo  {journal} {Optics Letters}\ }\textbf {\bibinfo {volume}
  {38}},\ \bibinfo {pages} {4605} (\bibinfo {year} {2013})}\BibitemShut
  {NoStop}%
\bibitem [{\citenamefont {Guizar-Sicairos}\ \emph {et~al.}(2008)\citenamefont
  {Guizar-Sicairos}, \citenamefont {Thurman},\ and\ \citenamefont
  {Fienup}}]{guizar2008efficient}%
  \BibitemOpen
  \bibfield  {author} {\bibinfo {author} {\bibfnamefont {M.}~\bibnamefont
  {Guizar-Sicairos}}, \bibinfo {author} {\bibfnamefont {S.~T.}\ \bibnamefont
  {Thurman}}, \ and\ \bibinfo {author} {\bibfnamefont {J.~R.}\ \bibnamefont
  {Fienup}},\ }\href@noop {} {\bibfield  {journal} {\bibinfo  {journal} {Optics
  letters}\ }\textbf {\bibinfo {volume} {33}},\ \bibinfo {pages} {156}
  (\bibinfo {year} {2008})}\BibitemShut {NoStop}%
\bibitem [{\citenamefont {Paganin}(2006)}]{paganin2006coherent}%
  \BibitemOpen
  \bibfield  {author} {\bibinfo {author} {\bibfnamefont {D.}~\bibnamefont
  {Paganin}},\ }\href@noop {} {\emph {\bibinfo {title} {Coherent X-ray
  optics}}},\ \bibinfo {number} {6}\ (\bibinfo  {publisher} {Oxford University
  Press on Demand},\ \bibinfo {year} {2006})\BibitemShut {NoStop}%
\bibitem [{\citenamefont {Als-Nielsen}\ and\ \citenamefont
  {McMorrow}(2011)}]{als2011}%
  \BibitemOpen
  \bibfield  {author} {\bibinfo {author} {\bibfnamefont {J.}~\bibnamefont
  {Als-Nielsen}}\ and\ \bibinfo {author} {\bibfnamefont {D.}~\bibnamefont
  {McMorrow}},\ }\href@noop {} {\emph {\bibinfo {title} {Elements of modern
  X-ray physics}}}\ (\bibinfo  {publisher} {John Wiley \& Sons},\ \bibinfo
  {year} {2011})\BibitemShut {NoStop}%
\bibitem [{\citenamefont {Morgan}\ \emph {et~al.}(2016)\citenamefont {Morgan},
  \citenamefont {Petersen}, \citenamefont {Donnelley}, \citenamefont {Farrow},
  \citenamefont {Parsons},\ and\ \citenamefont {Paganin}}]{morgan2016}%
  \BibitemOpen
  \bibfield  {author} {\bibinfo {author} {\bibfnamefont {K.~S.}\ \bibnamefont
  {Morgan}}, \bibinfo {author} {\bibfnamefont {T.~C.}\ \bibnamefont
  {Petersen}}, \bibinfo {author} {\bibfnamefont {M.}~\bibnamefont {Donnelley}},
  \bibinfo {author} {\bibfnamefont {N.}~\bibnamefont {Farrow}}, \bibinfo
  {author} {\bibfnamefont {D.~W.}\ \bibnamefont {Parsons}}, \ and\ \bibinfo
  {author} {\bibfnamefont {D.~M.}\ \bibnamefont {Paganin}},\ }\href@noop {}
  {\bibfield  {journal} {\bibinfo  {journal} {Optics Express}\ }\textbf
  {\bibinfo {volume} {24}},\ \bibinfo {pages} {24435} (\bibinfo {year}
  {2016})}\BibitemShut {NoStop}%
\bibitem [{\citenamefont {Kitchen}\ \emph {et~al.}(2008)\citenamefont
  {Kitchen}, \citenamefont {Lewis}, \citenamefont {Morgan}, \citenamefont
  {Wallace}, \citenamefont {Siew}, \citenamefont {Siu}, \citenamefont {Habib},
  \citenamefont {Fouras}, \citenamefont {Yagi}, \citenamefont {Uesugi} \emph
  {et~al.}}]{kitchen2008}%
  \BibitemOpen
  \bibfield  {author} {\bibinfo {author} {\bibfnamefont {M.~J.}\ \bibnamefont
  {Kitchen}}, \bibinfo {author} {\bibfnamefont {R.}~\bibnamefont {Lewis}},
  \bibinfo {author} {\bibfnamefont {M.~J.}\ \bibnamefont {Morgan}}, \bibinfo
  {author} {\bibfnamefont {M.~J.}\ \bibnamefont {Wallace}}, \bibinfo {author}
  {\bibfnamefont {M.}~\bibnamefont {Siew}}, \bibinfo {author} {\bibfnamefont
  {K.~K.~W.}\ \bibnamefont {Siu}}, \bibinfo {author} {\bibfnamefont
  {A.}~\bibnamefont {Habib}}, \bibinfo {author} {\bibfnamefont
  {A.}~\bibnamefont {Fouras}}, \bibinfo {author} {\bibfnamefont
  {N.}~\bibnamefont {Yagi}}, \bibinfo {author} {\bibfnamefont {K.}~\bibnamefont
  {Uesugi}},  \emph {et~al.},\ }\href@noop {} {\bibfield  {journal} {\bibinfo
  {journal} {Physics in Medicine \& Biology}\ }\textbf {\bibinfo {volume}
  {53}},\ \bibinfo {pages} {6065} (\bibinfo {year} {2008})}\BibitemShut
  {NoStop}%
\bibitem [{\citenamefont {Zentai}(2008)}]{zentai2008}%
  \BibitemOpen
  \bibfield  {author} {\bibinfo {author} {\bibfnamefont {G.}~\bibnamefont
  {Zentai}},\ }in\ \href@noop {} {\emph {\bibinfo {booktitle} {2008 IEEE
  International Workshop on Imaging Systems and Techniques}}}\ (\bibinfo
  {organization} {IEEE},\ \bibinfo {year} {2008})\ pp.\ \bibinfo {pages}
  {1--6}\BibitemShut {NoStop}%
\bibitem [{\citenamefont {Miller}\ \emph {et~al.}(2013)\citenamefont {Miller},
  \citenamefont {White}, \citenamefont {McDonald},\ and\ \citenamefont
  {Seifert}}]{miller2013}%
  \BibitemOpen
  \bibfield  {author} {\bibinfo {author} {\bibfnamefont {E.~A.}\ \bibnamefont
  {Miller}}, \bibinfo {author} {\bibfnamefont {T.~A.}\ \bibnamefont {White}},
  \bibinfo {author} {\bibfnamefont {B.~S.}\ \bibnamefont {McDonald}}, \ and\
  \bibinfo {author} {\bibfnamefont {A.}~\bibnamefont {Seifert}},\ }\href@noop
  {} {\bibfield  {journal} {\bibinfo  {journal} {IEEE Transactions on Nuclear
  Science}\ }\textbf {\bibinfo {volume} {60}},\ \bibinfo {pages} {416}
  (\bibinfo {year} {2013})}\BibitemShut {NoStop}%
\end{thebibliography}%


\end{document}